# $p$-WAVE MESONS EMITTING WEAK DECAYS OF BOTTOM MESONS

Maninder Kaur [a], Supreet Pal Singh[b] and R. C. Verma [c]
*Department of Physics, Punjabi University,*
*Patiala – 147002, India.*

Abstract

This paper is the extension of our previous work entitled "Searching a systematics for nonfactorizable contributions to $B^-$ and $\overline{B}^0$ hadronic decays". Obtaining the factorizable contributions from the spectator-quark model for $N_c = 3,$ a systematics has been identified among the isospin reduced amplitudes for the nonfactorizable terms among $\overline{B} \to D\pi / D^*\pi / D\rho$ decay modes. This systematics helps us to derive a generic formula which assists to predict the branching fractions for $\overline{B}^0 -$decays. Inspired by this observation, we extend our analysis to $p$-wave meson emitting decays of $B-$meson $\overline{B} \to PA / PT / PS,$ particularly $\overline{B} \to a_1 D / \pi D_1 / \pi D'_1 / \pi D_2 / \pi D_0$, which have similar isospin structure and make predictions for $\overline{B}^0 -$decays, where the experimental measurements are not yet available.

[a] maninderphy@gmail.com
[b] spsmudahar@gmail.com
[c] rcverma@gmail.com

# I. INTRODUCTION

At present, large amount of information on the decays of the heavy flavor hadrons is available, and more measurements are expected in the future experiments. All over the world, several groups at Fermi lab, Cornell, LHC-CERN, KEK, DESY and Beijing Electron Collider etc., have been working to provide wide knowledge of the heavy flavor physics. One of the goals of heavy flavor hadron physics is also to elucidate the relationship among the particles of different generations [1].

Being heavy, charm and bottom mesons have revealed many channels for leptonic, semi-leptonic and hadronic decays. The $b$ quark is especially interesting in this respect, as it has $W$-mediated transitions to both first generation (u) and second generation (c) quarks. The Standard Model is reasonably successful in explaining the leptonic and semileptonic decays, but the issue of weak hadronic decays is yet to be settled, and these decays have posed serious problems due to the strong interaction interference with the weak interactions responsible for these decays [2-8]. Initially one expected the weak hadronic decays of charm and bottom mesons to have less interference due to the strong interactions, as their decay products carry large momenta. However, their measurements have revealed the contrary. The prominent reason being that experiments are producing data at the hadronic level, whereas the theory (SM) deals with quarks and leptons. Presently the problem of Hadronization (formation of hadrons from quarks), being low energy phenomenon, cannot yet be resolved from the first principles. In fact, understanding of the decays gets more complicated as the produced hadrons in the weak hadronic decays can participate in the Final State Interactions (FSI) [9-20] caused by the strong interactions at hadronic level. Therefore, analysis of weak hadronic decays require phenomenological treatment for which symmetry principles and quark models are often employed for exploring the dynamics involved.

Even the weak interaction vertex itself is also affected through gluons-exchange among the quarks involved. At $W$-mass scale, hard gluons exchange effects are calculable using the perturbative QCD. Usually, factorization of weak matrix elements is performed in terms of certain form-factors and decay constants. Besides these high energy gluon exchanges, there exist possibility of soft gluon exchanges around the $W$- vertex, which generate the nonfactorizable contributions in the weak matrix elements [21-24]. The nonfactorizable terms may appear for several reasons, like soft gluon exchange, FSI rescattering effects [21-24]. The rescattering effects on the outgoing mesons have been studied in detail for bottom meson decays [25-26]. Besides that, flavor $SU(3)$ symmetry and Factorization Assisted Topological (FAT) approach have been employed for the study of such nonfactorizable contributions, as they have the advantage of absorbing various kinds of contributions lump- sum in terms of a few parameters, to be fixed empirically [27-28]. Extensive work has also been conducted to treat nonfactorizable contributions such as QCD factorization approach based on collinear factorization theorem [29] and the perturbative QCD factorization approach [30-31]. Unfortunately, it is not straightforward to calculate such terms, which are nonperturbative in nature, and require empirical data to investigate their behaviour.

In the naïve factorization scheme, the nonfactorizable contribution for the decay amplitudes was totally ignored and the two QCD coefficients $a_1$ and $a_2$ are fixed from the experimental data [32-34]. Initially, data on branching fractions of $D \to \bar{K}\pi$ decays seemed to require, $a_1 \approx c_1 = 1.26$, $a_2 \approx c_2 = -0.51$, leading to destructive interference between color-favored (CF) and color-suppressed (CS) processes for $D^+ \to \bar{K}^0\pi^+$, thereby implying the $N_c \to \infty$ limit [35]. However, later the measurement of $\bar{B} \to D\pi$ meson decays did not favor this result empirically, as these decays require $a_1 \approx 1.03$, and $a_2 \approx 0.23$, *i.e.* a positive value of $a_2$, in sharp contrast to the expectations based on the large $N_c$ limit, because the final state particles leave the interaction region very quickly allowing little time for final state interactions and soft-gluon exchange becomes less important [36]. Thus *B*- meson decays, revealing constructive interference between CF and CS diagrams for $B^- \to \pi^- D^0$, seem to favor $N_c = 3$ (real value).

It has been found experimentally that two-body decays dominate the spectrum. Bottom meson decays to two *s*- wave mesons (pseudoscalar and vector mesons) have been studied reasonably well. Theoretical focus has also, so far, been on the *s*-wave meson ($\bar{B} \to PP / PV$), emitting decays [1-25]. There exist four L=1 states: scalar ($J^{PC} = 0^{++}$), axial-vectors ($J^{PC} = 1^{++}$ and $1^{+-}$) and tensor ($J^{PC} = 2^{++}$) mesons. All these *p*-wave states and vector mesons decays to *s*-wave mesons through strong interactions, so these are called meson resonances. These states are generally produced either in the scattering experiments or as decay products of heavy flavour mesons. Thus, we investigate the following decay modes involving one *p*- wave meson in the final state:

$$\bar{B} \to P(0^{-+}) + S(0^{++}),$$
$$\bar{B} \to P(0^{-+}) + A(1^{++}),$$
$$\bar{B} \to P(0^{-+}) + A'(1^{+-}),$$
$$\bar{B} \to P(0^{-+}) + T(2^{++}).$$

Branching fractions for some of the decay modes have been measured experimentally [1]. Kinematically, these decays are expected to be suppressed, however, the measured branching fractions of these modes are rather large. Therefore, it is desirable to study *B*- meson decays emitting *p*- wave mesons, which requires theoretical understanding.

Our group performed a thorough study of nonfactorizable contributions by using isospin analysis for $D \to \bar{K}\pi / \bar{K}\rho / \bar{K}^*\pi$ decay modes and recognized a systematics for the ratio of nonfactorized reduced amplitudes. It is worth pointing out that this systematics was also found to be consistent with *p*-wave meson emitting decays of charm mesons: $D \to \bar{K}a_1 / \pi\bar{K}_1 / \pi\bar{\underline{K}}_1 / \pi\bar{K}_0 / \bar{K}a_2$ [23]. In our previous work [37], it has been found that the nonfactorizable contributions in the respective 1/2 and 3/2 isospin reduced amplitudes for Cabibo-favored $\bar{B} \to \pi D / \rho D / \pi D^*$ decay modes bear universal ratio equal to $\alpha$ within the experimental errors. Extension of this universality to $\bar{B} \to a_1 D / \pi D_1 / \pi D'_1 / \pi D_2 / \pi D_0$ is hoped to yield useful predictions for their branching fractions. Therefore, in this paper, we extend our analysis to investigate nonfactorizable terms in the *p*-wave mesons emitting decays.

This paper is organised as follow. In section II, weak Hamiltonian is expressed as sum of two particle generating factorizable and nonfactorizable contributions to the hadronic decays of *B*-mesons. In section III, we illustrate brief methodology of our approach by analysing *s*-wave mesons emitting decays of bottom mesons. In Session IV, detailed analysis of *p*-wave meson emitting decays is presented. Summary and conclusions are given in the last section.

## II. WEAK HAMILTONIAN

In order to study the two body hadronic $B-$decays, we consider the effective weak Hamiltonian [38]

$$H_w = \frac{G_F}{\sqrt{2}} V_{cb} V_{ud}^* \left[ c_1 \left(\bar{d}u\right)\left(\bar{c}b\right) + c_2 \left(\bar{c}u\right)\left(\bar{d}b\right) \right], \tag{1}$$

where $V_{ud}$ and $V_{cb}$ are the Cabibbo–Kobayashi–Maskawa (CKM) matrix elements [1],

$$V_{ud} = 0.975, \qquad V_{cb} = 0.041,$$

$\bar{q}_1 q_2 = \bar{q}_1 \gamma_\mu \left(1-\gamma_5\right) q_2$ denotes color singlet *V–A* Dirac current and the QCD coefficients at the bottom mass scale are taken as [39-40]

$$c_1 = 1.132, \qquad c_2 = -0.287. \tag{2}$$

In the standard factorization scheme, the current operators in the weak Hamiltonian are expressed in terms of the fundamental quark fields, it is appropriate to have the Hamiltonian in a form such that one of these currents carries the same quantum numbers as one of the mesons emitted in the final state of bottom meson decays. Consequently, the hadronic matrix elements of the Hamiltonian operator $H_w$ receives contribution from the operator itself and from its Fierz transformation. For instance, separating the factorizable and nonfactorizable parts of $(\bar{d}u)(\bar{c}b)$ using the Fierz identity [41], as given

$$(\bar{d}u)(\bar{c}b) = \frac{1}{N_c}(\bar{c}u)(\bar{d}b) + \frac{1}{2}\left(\bar{c}\lambda^a u\right)\left(\bar{d}\lambda^a b\right), \tag{3}$$

where $\bar{q}_1 \lambda^a q_2 \equiv \bar{q}_1 \gamma_\mu \left(1-\gamma_5\right) \lambda^a q_2$ represents the color octet current and performing similar treatment on other operator $(\bar{c}u)(\bar{d}b)$, the weak Hamiltonian finally becomes

$$H_w^{CF} = \frac{G_F}{\sqrt{2}} V_{cb} V_{ud}^* \left[ a_1 \left(\bar{d}u\right)_H \left(\bar{c}b\right)_H + c_2 H_w^8 \right], \tag{4}$$

$$H_w^{CS} = \frac{G_F}{\sqrt{2}} V_{cb} V_{ud}^* \left[ a_2 \left(\bar{c}u\right)_H \left(\bar{d}b\right)_H + c_1 \tilde{H}_w^8 \right], \tag{5}$$

for the color-favored (CF) and color-suppressed (CS) processes, respectively, where

$$a_{1,2} = c_{1,2} + \frac{c_{2,1}}{N_c}, \tag{6}$$

$$H_w^8 = \frac{1}{2}\sum_{a=1}^{8}\left(\bar{c}\lambda^a u\right)\left(\bar{d}\lambda^a b\right), \qquad \tilde{H}_w^8 = \frac{1}{2}\sum_{a=1}^{8}\left(\bar{d}\lambda^a u\right)\left(\bar{c}\lambda^a b\right). \tag{7}$$

Here the subscript $H$ in (4) and (5) indicates the change from quark current to hadron field operator [4]. Matrix elements of the first terms in (4) and (5) lead to the factorizable contributions [4] and the second terms, involving the color octet currents, generate nonfactorizable contributions.

## III. ANALYSIS OF *s*- WAVE MESON EMITTING DECAYS OF $\bar{B}$-MESONS

In this section, we illustrate methodology of our approach by analysing $\bar{B} \to PP$, decay mode. The branching fraction for $B$-mesons decay into two pseudoscalar mesons is related to its decay amplitude as follows:

$$B\left(\bar{B} \to P_1 P_2\right) = \tau_B \left|\frac{G_F}{\sqrt{2}} V_{cb} V_{ud}^*\right|^2 \frac{p}{8\pi m_B^2}\left|A\left(\bar{B} \to P_1 P_2\right)\right|^2, \tag{8}$$

where $\tau_B$ denotes the lifetime of $B$-mesons measured to be [1],

$$\tau_{\bar{B}^0} = (1.519 \pm 0.004) \times 10^{-12} \text{ s}, \qquad \tau_{B^-} = (1.638 \pm 0.004) \times 10^{-12} \text{ s},$$

and $p$ is the magnitude of the 3-momentum of the final state particles in the rest frame of parent $B$- meson,

$$p = |p_1| = |p_2| = \frac{1}{2m_B}\left[\left\{m_B^2 - \left(m_1 + m_2\right)^2\right\}\left\{m_B^2 - \left(m_1 - m_2\right)^2\right\}\right]^{1/2}.$$

Using the isospin framework, $\bar{B} \to \pi D$ decay amplitudes are represented in terms of isospin reduced amplitudes ($A_{1/2}^{\pi D}$, $A_{3/2}^{\pi D}$), and the strong interaction phases ($\delta_{1/2}^{\pi D}$, $\delta_{3/2}^{\pi D}$) in respective Isospin -1/2 and 3/2 final states, as

$$\begin{aligned} A(\bar{B}^0 \to \pi^- D^+) &= \frac{1}{\sqrt{3}}\left[A_{3/2}^{\pi D} e^{i\delta_{3/2}^{\pi D}} + \sqrt{2} A_{1/2}^{\pi D} e^{i\delta_{1/2}^{\pi D}}\right], \\ A(\bar{B}^0 \to \pi^0 D^0) &= \frac{1}{\sqrt{3}}\left[\sqrt{2} A_{3/2}^{\pi D} e^{i\delta_{3/2}^{\pi D}} - A_{1/2}^{\pi D} e^{i\delta_{1/2}^{\pi D}}\right], \\ A(B^- \to \pi^- D^0) &= \sqrt{3} A_{3/2}^{\pi D} e^{i\delta_{3/2}^{\pi D}}. \end{aligned} \tag{9}$$

These equations lead to the following relations:

$$A_{1/2}^{\pi D} = \left[ \left| A(\bar{B}^0 \to \pi^- D^+) \right|^2 + \left| A(\bar{B}^0 \to \pi^0 D^0) \right|^2 - \frac{1}{3} \left| A(B^- \to \pi^- D^0) \right|^2 \right]^{1/2},$$

$$A_{3/2}^{\pi D} = \sqrt{\frac{1}{3}} \left| A(B^- \to \pi^- D^0) \right|, \tag{10}$$

The experimental values [1],

$$B\left(\bar{B}^0 \to \pi^- D^+\right) = \left(2.52 \pm 0.13\right) \times 10^{-3},$$
$$B\left(\bar{B}^0 \to \pi^0 D^0\right) = \left(2.63 \pm 0.14\right) \times 10^{-4},$$
$$B\left(B^- \to \pi^- D^0\right) = \left(4.68 \pm 0.13\right) \times 10^{-3},$$

yields

$$A_{1/2}^{\pi D\ \exp} = \pm\left(1.273 \pm 0.065\right) \text{GeV}^3, \qquad A_{3/2}^{\pi D\ \exp} = \pm\left(1.323 \pm 0.018\right) \text{ GeV}^3. \tag{11}$$

We express decay amplitude as sum of the factorizable and nonfactorizable parts,

$$A(\bar{B} \to \pi D) = A^f(\bar{B} \to \pi D) + A^{nf}(\bar{B} \to \pi D), \tag{12}$$

arising from the respective terms of the weak Hamiltonian given in (4) and (5).

Using the factorization scheme, spectator-quark parts of the decay amplitudes arising from *W*- emission diagrams are derived for the following classes of $\bar{B} \to \pi D$ decays.

$$A^f(\bar{B}^0 \to \pi^- D^+) = a_1 f_\pi \left(m_B^2 - m_D^2\right) F_0^{\bar{B}D}\left(m_\pi^2\right) = \left(2.180 \pm 0.099\right) \text{ GeV}^3, \tag{13}$$

$$A^f(\bar{B}^0 \to \pi^0 D^0) = -\frac{1}{\sqrt{2}} a_2 f_D \left(m_B^2 - m_\pi^2\right) F_0^{\bar{B}\pi}\left(m_D^2\right) = -\left(0.111 \pm 0.021\right) \text{GeV}^3, \tag{14}$$

$$A^f(B^- \to \pi^- D^0) = a_1 f_\pi \left(m_B^2 - m_D^2\right) F_0^{\bar{B}D}\left(m_\pi^2\right) + a_2 f_D \left(m_B^2 - m_\pi^2\right) F_0^{\bar{B}\pi}\left(m_D^2\right)$$
$$= \left(2.339 \pm 0.103\right) \text{GeV}^3. \tag{15}$$

Numerical inputs for decay constant,

$$f_D = \left(0.207 \pm 0.009\right) \text{GeV}, \quad f_\pi = \left(0.131 \pm 0.002\right) \text{ GeV}, \tag{16}$$

taken from the leptonic decays of $D$ and $\pi$ mesons, respectively [42].

Assuming the nearest pole dominance, momentum dependence of the form- factors, appearing in the decay amplitudes given in (13-15), is taken as

$$F_0\left(q^2\right) = \frac{F_0(0)}{\left(1 - q^2 / m_s^2\right)}, \tag{17}$$

where the pole masses $m_s$ are given by the lowest lying meson with the appropriate quantum numbers i.e. $J^P = 0^+$ for $F_0(0)$ and $1^-$ for $F_1(0)$. For numerical estimation, we have taken scalar meson carrying the quantum number of the corresponding weak currents, which are $m_s(0^+) = 5.78$ GeV, and $m_s(0^+) = 6.80$ GeV [2-4, 22, 47, 43]. Form-factors $F_0(0)$ at $q^2 = 0$ are taken from [44], as given below,

$$F_0^{\bar{B}\pi}(0) = (0.27 \pm 0.05), \quad F_0^{\bar{B}D}(0) = (0.66 \pm 0.03). \tag{18}$$

Exploiting the following isospin relations:

$$\begin{aligned} A_{1/2}^f(\bar{B} \to \pi D) &= \frac{1}{\sqrt{3}}\left\{\sqrt{2}A^f(\bar{B}^0 \to \pi^- D^+) - A^f(\bar{B}^0 \to \pi^0 D^0)\right\}, \\ A_{3/2}^f(\bar{B} \to \pi D) &= \frac{1}{\sqrt{3}}\left\{A^f(\bar{B}^0 \to \pi^- D^+) + \sqrt{2}A^f(\bar{B}^0 \to \pi^0 D^0)\right\}, \end{aligned} \tag{19}$$

we obtain

$$A_{1/2}^f = (1.845 \pm 0.082) \text{ GeV}^3, \qquad A_{3/2}^f = (1.168 \pm 0.060) \text{ GeV}^3. \tag{20}$$

We write the non-factorizable part of the decay amplitudes in terms of isospin CG coefficients [22-23] as scattering amplitudes for spurion $+ \bar{B} \to \pi D$ process:

$$\begin{aligned} A^{nf}(\bar{B}^0 \to \pi^- D^+) &= \frac{1}{3}c_2\left(\left\langle \pi D \left\| H_w^8 \right\| \bar{B}\right\rangle_{3/2} + 2\left\langle \pi D \left\| H_w^8 \right\| \bar{B}\right\rangle_{1/2}\right), \\ A^{nf}(\bar{B}^0 \to \pi^0 D^0) &= \frac{\sqrt{2}}{3}c_1\left(\left\langle \pi D \left\| \tilde{H}_w^8 \right\| \bar{B}\right\rangle_{3/2} - \left\langle \pi D \left\| \tilde{H}_w^8 \right\| \bar{B}\right\rangle_{1/2}\right), \\ A^{nf}(B^- \to \pi^- D^0) &= c_2\left\langle \pi D \left\| H_w^8 \right\| \bar{B}\right\rangle_{3/2} + c_1\left\langle \pi D \left\| \tilde{H}_w^8 \right\| \bar{B}\right\rangle_{3/2}, \end{aligned} \tag{21}$$

where spurion is the fictitious particle carrying the quantum numbers of the weak Hamiltonian. At present, there is no available technique to calculate these quantities exactly from the theory of strong interactions. Therefore, subtracting the factorizable part (20) from the experimental decay amplitudes (11), we determine the nonfactorizable isospin reduced amplitudes,

$$A_{1/2}^{nf} = -(0.572 \pm 0.105)\,\text{GeV}^3, \quad A_{3/2}^{nf} = -(2.491 \pm 0.062)\,\text{GeV}^3. \tag{22}$$

by choosing positive and negative values for $A_{1/2}^{\pi D\ \exp}$ and $A_{3/2}^{\pi D\ \exp}$, respectively, from (11). These bear the following ratio:

$$\alpha = \left(\frac{A_{1/2}^{nf}}{A_{3/2}^{nf}}\right)_{\bar{B} \to \pi D} = 0.229 \pm 0.042. \tag{23}$$

Such isospin formalism can easily be extended to $\bar{B} \to \rho D$ and $\bar{B} \to \pi D^*$ decays, as the isospin structure of these decay modes is exactly the same as that of $\bar{B} \to \pi D$. Following the procedure discussed for the $\bar{B} \to \pi D$ mode, we calculate the ratio of non-factorizable isospin parts for $\bar{B} \to \rho D$ and $\bar{B} \to \pi D^*$ decay modes, shown below for the sake of comparison:

$$\begin{array}{ccccc} \dfrac{A_{1/2}^{nf}(\bar{B} \to \rho D)}{A_{3/2}^{nf}(\bar{B} \to \rho D)} & \simeq & \dfrac{A_{1/2}^{nf}(\bar{B} \to \pi D^*)}{A_{3/2}^{nf}(\bar{B} \to \pi D^*)} & \simeq & \dfrac{A_{1/2}^{nf}(\bar{B} \to \pi D)}{A_{3/2}^{nf}(\bar{B} \to \pi D)}. \\ 0.200 \pm 0.096 & & 0.211 \pm 0.109 & & 0.229 \pm 0.042 \end{array} \tag{24}$$

Note that, the ratios of $\alpha = \dfrac{A_{1/2}^{nf}}{A_{3/2}^{nf}}$ for all the three decay modes $\bar{B} \to \pi D / \rho D / \pi D^*$ are in consistent agreement with each other.

In fact, these relations can be expressed in a generic form as

$$B_{-+} + B_{00} = \frac{\tau_{\bar{B}^0}}{3\tau_{B^-}} B_{0-} \left[ 1 + \left\{ \alpha + \frac{\left(\sqrt{2} - \alpha\right) A_{-+}^f - \left(1 + \sqrt{2}\alpha\right) A_{00}^f}{A_{0-}} \right\}^2 \right], \tag{25}$$

with $\alpha = \dfrac{A_{1/2}^{nf}}{A_{3/2}^{nf}}$, where subscripts of branching fractions, $B_{-+}, B_{00}, B_{0-}$, denote the charge states of the non-charm and charm mesons respectively emitted in each case. $A_{-+}^f$ and $A_{00}^f$ factorizable amplitudes denote the charge state of the mesons for $\bar{B}^0-$decays. $A_{0-}$, the total decay amplitude, is obtained from the $B^-$ – decay as,

$$A_{0\text{-}} = \sqrt{\frac{B_{0\text{-}}}{\tau_{B^-} \times (kinematic\ factor)}}, \tag{26}$$

where the kinematic factor for $\bar{B} \to PP$ and $\bar{B} \to PV$ are

$$\text{kinematic factor for } \bar{B} \to PP = \left| \frac{G_F}{\sqrt{2}} V_{cb} V_{ud}^* \right|^2 \frac{p}{8\pi m_B^2}, \tag{27}$$

$$\text{kinematic factor for } \bar{B} \to PV = \left| \frac{G_F}{\sqrt{2}} V_{cb} V_{ud}^* \right|^2 \frac{p^3}{8\pi m_V^2}, \tag{28}$$

and $p$ is the magnitude of the three-momentum of the final-state particle in the rest frame of $B$-meson and $m_B$ and $m_V$ denote masses of the $B$-meson and vector meson.

Taking the average value of $\alpha$ = 0.22, we have predicted sum of the branching fractions of $\bar{B}^0 -$ decays [37], as

$$B\left(\bar{B}^0 \to \pi^- D^+\right) + B\left(\bar{B}^0 \to \pi^0 D^0\right) = (0.28 \pm 0.02)\% \quad Theo, \\ = (0.28 \pm 0.01)\% \quad Expt; \tag{29}$$

$$B\left(\bar{B}^0 \to \rho^- D^+\right) + B\left(\bar{B}^0 \to \rho^0 D^0\right) = (0.76 \pm 0.13)\% \quad Theo, \\ = (0.79 \pm 0.12)\% \quad Expt; \tag{30}$$

$$B\left(\bar{B}^0 \to \pi^- D^{*+}\right) + B\left(\bar{B}^0 \to \pi^0 D^{*0}\right) = (0.29 \pm 0.04)\% \quad Theo, \\ = (0.30 \pm 0.01)\% \quad Expt. \tag{31}$$

All theoretical values match well within experimental errors.

## IV. ANALYSIS OF *p*- WAVE MESON EMITTING DECAYS OF $\bar{B}$ - MESONS

We, in this section, study the Cabibbo- favored *p*-wave meson emitting decays in the channels $\bar{B} \to PA / PT / PS$ involving $b + u \to c + d / s$ transitions. Naively, one may expect that these decays to be kinematically suppressed due to the large masses of the *p*-wave resonance. However, it has been found that their measured branching fractions compete well with that of the *s*-wave meson emitting decays of bottom mesons. On the experimental side, branching fractions of a few such decays have been measured as shown in the Table 1. Out of them $\bar{B} \to a_1 D$ decays have clean values for their branching fractions, whereas other branching fractions are measured in the composite form. From the results of *s*-wave analysis, we can extend this isospin formalism to $\bar{B} \to a_1 D / \pi D_1 / \pi D_1^{'} / \pi D_2 / \pi D_0$ decay modes, since the isospin structure of these decay modes is exactly the same as that of $\bar{B} \to \pi D$ mode.

**Table 1. Experimental Data for *p*-wave Meson Emitting Decays [1]**

| Channels | Branching Fraction of Decays | Experimental Branching Fractions [1] |
|---|---|---|
| $\bar{B} \to PA$ | $B(\bar{B}^0 \to a_1^- D^+)$ | $(6.0 \pm 3.3) \times 10^{-3}$ |
| | $B(B^- \to a_1^- D^0)$ | $(4 \pm 4) \times 10^{-3}$ |
| | $B(B^- \to \pi^- D_1(2.420)^0)$ | $(1.5 \pm 0.6) \times 10^{-3}$ |
| | $B(B^- \to \pi^- D_1^{'}(2.427)^0) \times B(D_1^{'}(2.427)^0 \to \pi^- D^{*+})$ | $(5.0 \pm 1.2) \times 10^{-4}$ |
| $\overline{B} \to PT$ | $B(B^- \to \pi^- D^*_2(2.462)^0) \times B(D^*_2(2.462)^0 \to \pi^- D^+)$<br>$B(B^- \to \pi^- D^*_2(2.462)^0) \times B(D^*_2(2.462)^0 \to \pi^- D^{*+})$ | $(3.56 \pm 0.24) \times 10^{-4}$<br>$(2.2 \pm 1.1) \times 10^{-4}$ |
| $\overline{B} \to PS$ | $B(B^- \to \pi^- D^*_0(2.400)^0) \times B(D^*_0(2.400)^0 \to \pi^- D^+)$ | $(6.4 \pm 1.4) \times 10^{-4}$ |

## A. $\overline{B} \to PA$ MODE

**Axial-Vector Meson Spectroscopy**

Experimentally [1], two types of the axial-vector mesons exist with different charge conjugation properties, *i.e.,* ${}^3P_1(J^{PC} = 1^{++})$ and ${}^1P_1(J^{PC} = 1^{+-})$, behave well with respect to the quark model $q\overline{q}$ assignments observed, strange and charmed states are given by mixture of ${}^3P_1$ and ${}^1P_1$ states. In contrast, hidden-flavor diagonal ${}^3P_1$ and ${}^1P_1$ states have opposite C-parity and therefore cannot mix. The following non-strange and uncharmed mesons have been observed (mass in GeV):

**1.** For ${}^3P_1$ multiplet,

i. Three isovector $a_1(1.230)$ with the quark content $u\overline{d}, u\overline{u} - d\overline{d}/\sqrt{2}$ and $d\overline{u}$, respectively, are

$$a_1^+, a_1^0 \text{ and } a_1^- . \qquad (32)$$

ii. Four isoscalars $f_1(1.285), f_1(1.420), f_1^{'}(1.512),$ and $\chi_{c1}(3.511)$ have been observed, out of which $f_1(1.420)$ is a multiquark state in the form of a $K\overline{K}\pi$ bound state [45] or a $K\overline{K}^*$ deuteron-state [46].

**2.** For ${}^1P_1$ multiplet,

i. Isovector $b_1(1.229)$ with flavor content same as given in (32) are

$$b_1^+, b_1^0 \text{ and } b_1^- . \qquad (33)$$

ii. Three isoscalars are $h_1(1.170), h_1^{'}(1.380)$ and $h_{c1}(3.526)$. C-parity of $h^{'}_1(1.380)$ and spin and parity of the $h_{c1}(3.526)$ remains to be confirmed.

Proximity of $a_1(1.230)$ and $f_1(1.285)$ and to a lesser extent that of $b_1(1.229)$ and $h_1(1.170)$ indicate the ideal mixing for both $1^{++}$ and $1^{+-}$ diagonal states.

The states involving a strange quark of $A(J^{PC}=1^{++})$ and $A^{'}(J^{PC}=1^{+-})$ multiplets mix to generate the physical states in the following manner [47-49]:

$$K_1(1.270) = K_{1A}\sin\theta_1 + K_{1A^{'}}\cos\theta_1,$$
$$\underline{K}_1(1.400) = K_{1A}\cos\theta_1 - K_{1A^{'}}\sin\theta_1, \tag{34}$$

where $K_{1A}$ and $K_{1A^{'}}$ denote the strange partners of $a_1(1.230)$ and $b_1(1.229)$ respectively. Particle Data Group [1] assumes that the mixing is maximal, *i.e.* $\theta_1 = 45^\circ$, whereas $\tau_1 \to K_1(1.270)/K_1(1.400)+\nu_\tau$ data yields $\theta_1 = \pm 37^\circ$ and $\theta_1 = \pm 58^\circ$ [50]. However, the study of $D \to K_1(1.270)\pi, K_1(1.400)\pi$ decays rules out positive mixing-angle solutions. As $D \to K_1^-(1.400)\pi^+$ gets largely suppressed for $\theta_1 = -37^\circ$ the solution $\theta_1 = -58^\circ$ [51] is experimentally favored.

The mixing of charmed ($c\bar{u}$ and $c\bar{d}$) and strange charmed ($c\bar{s}$) state mesons are given in similar fashion,

$$D_1(2.420) = D_{1A}\sin\theta_{D_1} + D_{1A^{'}}\cos\theta_{D_1},$$
$$D^{'}_1(2.427) = D_{1A}\cos\theta_{D_1} - D_{1A^{'}}\sin\theta_{D_1}, \tag{35}$$

and

$$D_{s1}(2.460) = D_{s1A}\sin\theta_{D_{s1}} + D_{s1A^{'}}\cos\theta_{D_{s1}},$$
$$\underline{D}_{s1}(2.535) = D_{s1A}\cos\theta_{D_{s1}} - D_{s1A^{'}}\sin\theta_{D_{s1}}. \tag{36}$$

However, in the heavy quark limit, the physical mass eigenstates with ($J^P = 1^+$) are $P_1^{3/2}$ and $P_1^{1/2}$ rather than ${}^3P_1$ and ${}^1P_1$ states, as the heavy quark spin $S_Q$ decouples from the other degrees of freedom, so that $S_Q$ and the total angular momentum of the light antiquark are separately, good quantum numbers. Therefore, heavy quark symmetry leads to

$$\left|P_1^{3/2}\right\rangle = \sqrt{\frac{2}{3}}\left|{}^1P_1\right\rangle + \sqrt{\frac{1}{3}}\left|{}^3P_1\right\rangle,$$
$$\left|P_1^{1/2}\right\rangle = \sqrt{\frac{1}{3}}\left|{}^1P_1\right\rangle - \sqrt{\frac{2}{3}}\left|{}^3P_1\right\rangle. \tag{37}$$

However, beyond the heavy quark limit, still there is a small mixing between $P_1^{3/2}$ and $P_1^{1/2}$ states denoted by

$$D_1(2.420) = D_1^{1/2}\cos\theta_2 + D_1^{3/2}\sin\theta_2,$$
$$D^{'}_1(2.427) = -D_1^{1/2}\sin\theta_2 + D_1^{3/2}\cos\theta_2, \tag{38}$$

Likewise for strange axial-vector charmed mesons,

$$D_{s1}(2.460) = D_{s1}^{1/2}\cos\theta_3 - D_{s1}^{3/2}\sin\theta_3,$$
$$D_{s1}(2.535) = D_{s1}^{1/2}\sin\theta_3 + D_{s1}^{3/2}\cos\theta_3, \qquad (39)$$

where the mixing angle $\theta_2 = -(5.7 \pm 2.4)^\circ$ is obtained by Belle Collaboration through a detailed $\overline{B} \to D^{*}\pi\,\pi$ analysis [52-53], while $\theta_3 \approx 7^\circ$ is obtained from the quark potential model [51]. We now consider $\overline{B} \to a_1 D$ and $\overline{B} \to \pi D_1 / \pi D_1^{'}$ decays in the following subsections.

### A(i) $\overline{B} \to a_1\, D$ Decay Mode

In this section, we illustrate methodology of our approach by analysing $\overline{B} \to PA,$ decay mode. The branching fraction for *B*-mesons decay into pseudoscalar and axial vector mesons is related to its decay amplitude as follows:

$$B\left(\overline{B} \to PA\right) = \tau_B \left|\frac{G_F}{\sqrt{2}} V_{cb} V_{ud}^{*}\right|^2 \frac{p^3}{8\pi m_A^2} \left|A\left(\overline{B} \to PA\right)\right|^2, \qquad (40)$$

and *p* is the magnitude of the 3-momentum of the final state particles in the rest frame of parent *B*- meson and $m_A$ denotes the mass of axial-vector meson.

$$p = |p_1| = |p_2| = \frac{1}{2m_B}\left[\left\{m_B^2 - \left(m_P + m_A\right)^2\right\}\left\{m_B^2 - \left(m_P - m_A\right)^2\right\}\right]^{1/2}.$$

Using the isospin framework, $\overline{B} \to a_1 D$ decay amplitudes are represented in terms of isospin reduced amplitudes ($A_{1/2}^{a_1 D}$, $A_{3/2}^{a_1 D}$), and the strong interaction phases ($\delta_{1/2}^{a_1 D}$, $\delta_{3/2}^{a_1 D}$) in respective Isospin -1/2 and 3/2 final states, as

$$A(\overline{B}^0 \to a_1^- D^+) = \frac{1}{\sqrt{3}}\left[A_{3/2}^{a_1 D} e^{i\delta_{3/2}^{a_1 D}} + \sqrt{2} A_{1/2}^{a_1 D} e^{i\delta_{1/2}^{a_1 D}}\right],$$
$$A(\overline{B}^0 \to a_1^0 D^0) = \frac{1}{\sqrt{3}}\left[\sqrt{2} A_{3/2}^{a_1 D} e^{i\delta_{3/2}^{a_1 D}} - A_{1/2}^{a_1 D} e^{i\delta_{1/2}^{a_1 D}}\right], \qquad (41)$$
$$A(B^- \to a_1^- D^0) = \sqrt{3} A_{3/2}^{a_1 D} e^{i\delta_{3/2}^{a_1 D}}.$$

These equations lead to the following relations:

$$A_{1/2}^{a_1 D} = \left[\left|A(\overline{B}^0 \to a_1^- D^+)\right|^2 + \left|A(\overline{B}^0 \to a_1^0 D^0)\right|^2 - \frac{1}{3}\left|A(B^- \to a_1^- D^0)\right|^2\right]^{1/2},$$
$$A_{3/2}^{a_1 D} = \sqrt{\frac{1}{3}}\left|A(B^- \to a_1^- D^0)\right|, \qquad (42)$$

and using the experimental value $B(B^- \to a_1^- D^0) = \left(4 \pm 4\right) \times 10^{-3}$, we get

$$A(B^- \to a_1^- D^0) = (0.25 \pm 0.25\,)\ \mathrm{GeV}^2. \qquad (43)$$

The isospin formalism assists us to derive a generic relation among the branching fractions of $\bar{B} \to a_1 D$ decays, as follow:

$$B(\bar{B}^0 \to a_1^- D^+) + B(\bar{B}^0 \to a_1^0 D^0) = \frac{\tau_{\bar{B}^0}}{3\tau_{B^-}} B(B^- \to a_1^- D^0) \times \left[ 1 + \left\{ \alpha + \frac{\left(\sqrt{2} - \alpha\right) A^f(\bar{B}^0 \to a_1^- D^+) - \left(1 + \sqrt{2}\alpha\right) A^f(\bar{B} \to a_1^0 D^0)}{A(B^- \to a_1^- D^0)} \right\}^2 \right], \tag{44}$$

with $\alpha \equiv \frac{A_{1/2}^{nf}}{A_{3/2}^{nf}} = 0.22$, from the analysis of *s*-wave meson emitting decays of $B-$mesons, inspired by the analysis of charm meson decays [23], where it has been observed that the *p*-wave mesons bears the same ratio as that of $D \to \pi \bar{K}$ decay mode.
Now we obtain factorizable amplitudes for $\bar{B}^0$ decays,

$$A^f(\bar{B}^0 \to a_1^- D^+) = 2a_1 m_{a_1} f_{a_1} F_1^{\bar{B}D}\left(m_{a_1}^2\right) = (0.369 \pm 0.016)\,\text{GeV}^2, \tag{45}$$

$$A^f(\bar{B}^0 \to a_1^0 D^0) = -\frac{1}{\sqrt{2}} a_2 m_{a_1} f_D V_0^{\bar{B}a_1}\left(m_D^2\right) = -(0.0033 \pm 0.0001)\,\text{GeV}^2, \tag{46}$$

where the decay constants are taken from [42],

$$f_{a_1} = -\left(0.203 \pm 0.018\right)\text{GeV}, \quad f_D = \left(0.207 \pm 0.009\right)\ \text{GeV}, \tag{47}$$

The form-factor $V_0^{Ba_1}(m_D^2)$ is obtained from CLFQM [42] results with the following $q^2$ dependence.

$$V_0^{\bar{B}a_1}\left(q^2\right) = \frac{V_0^{\bar{B}a_1}(0)}{\left(1 - a\left(\frac{q^2}{m_B^2}\right) - b\left(\frac{q^2}{m_B^2}\right)^2\right)}, \tag{48}$$

where

$$V_0^{\bar{B}a_1}\left(0\right) = 0.14 \pm 0.01, \quad a = 1.66 \pm 0.04, \quad b = 1.11 \pm 0.08, \tag{49}$$

and

$$F_1^{\bar{B}D}\left(0\right) = F_0^{\bar{B}D}\left(0\right) = (0.67 \pm 0.01),$$

which has already been used in (18). Finally, taking $B(B^- \to a_1^- D^0) = \left(4 \pm 4\right) \times 10^{-3}$ we predict

$$B(\bar{B}^0 \to a_1^- D^+) + B(\bar{B} \to a_1^0 D^0) = \begin{cases} \left(4.7 \pm 0.7\right) \times 10^{-3}\, for\ B(B^- \to a_1^- D^0) = 4 \times 10^{-3}, \\ \left(5.6 \pm 0.3\right) \times 10^{-3}\, for\ B(B^- \to a_1^- D^0) = 8 \times 10^{-3}, \end{cases} \tag{50}$$

which are barely touching the experimental value $B(\bar{B}^0 \to a_1^- D^+) = (6.4 \pm 3.3) \times 10^{-3}$.

There are several existing model calculations for the $\bar{B} \to A$ form factors: the ISGW2 model [5], the constituent quark–meson model (CQM) [54], the QSR [55], LCSR [56] and more recently the pQCD approach [57]. For the sake of comparison, results for the $\bar{B} \to a_1$ transition form factors are given in Table 2 for these approaches, which show quite significant differences since these approaches differ in their treatment of dynamics of the form-factors. Specially, $V_0^{\bar{B}a_1} = 1.20$ obtained in the quark–meson model and 1.01 in the ISGW2 model is larger than the values obtained in other approaches.

**Table 2. Form-factor of the $\bar{B} \to a_1$ transitions at maximum recoil ($q^2$ =0). The results of CQM and QSR have been rescaled according to the form-factor definition.**

| $\bar{B} \to a_1$ | CLFQM [42] | ISGW2 [3] | CQM [54] | QSR [55] | LCSR [56] | pQCD [57] |
|---|---|---|---|---|---|---|
| $V_0$ | 0.14±0.01 | 1.01 | 1.20 | 0.23±0.05 | 0.30±0.05 | 0.34±0.07 |

Looking at these uncertainties, in Fig 1, we present variation of the sum of

$$\sum B\left(\bar{B}^0 \to decays\right) \equiv B(\bar{B}^0 \to a_1^- D^+) + B(\bar{B} \to a_1^0 D^0)$$

w.r.t $B(B^- \to a_1^- D^0)$ for different values of form factor, $V_0^{\bar{B}a_1}(0) = 0.14$ and 1.20 (the dash line corresponds to $V_0^{\bar{B}a_1}(0) = 0.14$, and the solid line corresponds to $V_0^{\bar{B}a_1}(0) = 1.20$), which enhances our prediction by a factor of 1.19, *i.e.,*

$$B(\bar{B}^0 \to a_1^- D^+) + B(\bar{B} \to a_1^0 D^0) = (5.6 \pm 0.3) \times 10^{-3}.$$

We also notice that the present data favour $B(B^- \to a_1^- D^0)$ to be on the higher side. A new measurement of branching fractions of these decays would clarify the situation.

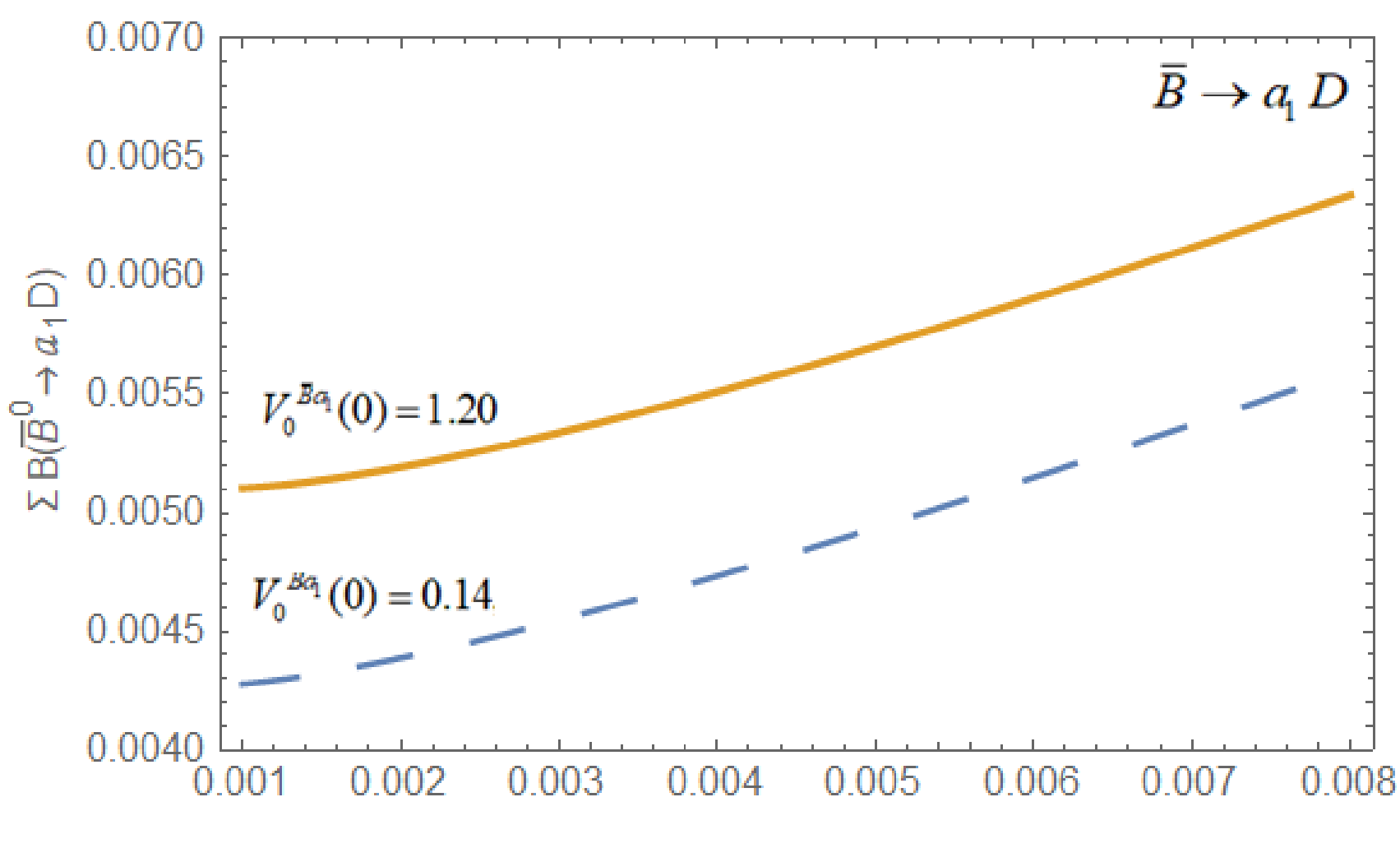

$B(B^- \to a_1^- D^0)$

**Fig 1. Variation of $\sum B\left(\bar{B}^0 \to a_1 D\right)$ with $B(B^- \to a_1^- D^0)$ for different values of form-factor**

### A(ii) $\bar{B} \to \pi D_1(2.420)$ Decay mode

We start with, writing the generic formula explicitly for $B^- \to \pi^- D_1(2.420)^0$

$$B(\bar{B}^0 \to \pi^- D_1^+) + B(\bar{B}^0 \to \pi^0 D_1^0)$$
$$= \frac{\tau_{\bar{B}^0}}{3\tau_{B^-}} B(B^- \to \pi^- D_1^0) \times \left[ 1 + \left\{ \alpha + \frac{\left(\sqrt{2} - \alpha\right) A^f(\bar{B}^0 \to \pi^- D_1^+) - \left(1 + \sqrt{2}\alpha\right) A^f(\bar{B}^0 \to \pi^0 D_1^0)}{A(B^- \to \pi^- D_1^0)} \right\}^2 \right]. \tag{51}$$

We take $\alpha \equiv \frac{A_{1/2}^{nf}}{A_{3/2}^{nf}} = 0.22$ from the analysis of *s*-wave meson emitting decays of $B-$mesons (23).

The experimental branching $B(B^- \to \pi^- D_1^0) = \left(1.5 \pm 0.6\right) \times 10^{-3}$, we get

$$A(B^- \to \pi^- D_1^0) = (0.213 \pm 0.040)\ \mathrm{GeV}^2, \tag{52}$$

Now we obtain factorizable amplitudes for $\bar{B}^0-$decays,

$$A^f(\bar{B}^0 \to \pi^- D_1^{\ +}) = 2a_1 m_{D_1} f_\pi V_0^{\bar{B}D_1}\left(m_\pi^2\right) = 0.332\,\text{GeV}^2,$$
$$A^f(\bar{B}^0 \to \pi^0 D_1^{\ 0}) = -\frac{1}{\sqrt{2}} 2a_2 m_{D_1} f_{D_1} F_1^{\bar{B}\pi}\left(m_{D_1}^2\right) = 0.012\ \text{GeV}^2, \quad (53)$$

where the decay constants are given by,

$$f_{D_1} = f_{D_1^{1/2}} \cos\theta_1 + f_{D_1^{3/2}} \sin\theta_1, \quad (54)$$

and

$$f_{D_1^{1/2}} = (0.179 \pm 0.035)\ \text{GeV},\ f_{D_1^{3/2}} = -(0.054 \pm 0.013)\ \text{GeV},$$
$$f_\pi = (0.131 \pm 0.002)\ \text{GeV}, \quad (55)$$

are taken from [42]. Required $\bar{B} \to D_1$ form factor is given by

$$V_0^{\bar{B}D_1}\left(m_\pi^2\right) = V_0^{\bar{B}D_1^{1/2}}\left(m_\pi^2\right)\cos\theta_1 + V_0^{\bar{B}D_1^{3/2}}\left(m_\pi^2\right)\sin\theta_1. \quad (56)$$

The form-factor $V_0^{\bar{B}D_1^{1/2}}(m_\pi^2)$ and $V_0^{\bar{B}D_1^{3/2}}(m_\pi^2)$ are taken from CLFQM [42] results with the following $q^2$ dependence.

$$V_0^{\bar{B}D_1^{1/2}}\left(q^2\right) = \frac{V_0^{\bar{B}D_1^{1/2}}(0)}{\left(1 - a\left(\frac{q^2}{m_B^2}\right) - b\left(\frac{q^2}{m_B^2}\right)^2\right)}, \quad (57)$$

$$V_0^{\bar{B}D_1^{3/2}}\left(q^2\right) = \frac{V_0^{\bar{B}D_1^{3/2}}(0)}{\left(1 - a\left(\frac{q^2}{m_B^2}\right) - b\left(\frac{q^2}{m_B^2}\right)^2\right)}, \quad (58)$$

where

$$V_0^{\bar{B}D_1^{1/2}}(0) = 0.11 \pm 0.01,$$
$$a = 1.08 \pm 0.02, \quad b = 0.08 \pm 0.03; \quad (59)$$

$$V_0^{\bar{B}D_1^{3/2}}(0) = 0.52 \pm 0.01,$$
$$a = 1.14 \pm 0.04, \quad b = 0.34 \pm 0.02. \quad (60)$$

The $\bar{B} \to \pi$ form factor,

$$F_1^{\bar{B}\pi}(0) = F_0^{\bar{B}\pi}(0) = (0.27 \pm 0.05),$$

has already been used in (18). For the charm mesons mixing angle $\theta_1 = -(5.7 \pm 2.4)^\circ$ we predict,

$$\sum B\left(\bar{B}^0 \to \pi D_1\right) \equiv B(\bar{B}^0 \to \pi^- D_1^{\ +}) + B(\bar{B}^0 \to \pi^0 D_1^{\ 0}) = \begin{cases} (4.7 \pm 1.7) \times 10^{-4} \ for\ \theta_1 = -8.1^\circ \\ (4.9 \pm 1.7) \times 10^{-4} \ for\ \theta_1 = -3.3^\circ \end{cases} \tag{61}$$

Here also, we plot variation of the sum of $B(\bar{B}^0 \to \pi^- D_1^{\ +})$ and $B(\bar{B}^0 \to \pi^0 D_1^{\ 0})$ w.r.t $B(B^- \to \pi^- D_1^{\ 0})$ in Fig 2, in the light of experimental error in $B(B^- \to \pi^- D_1^{\ 0})$.

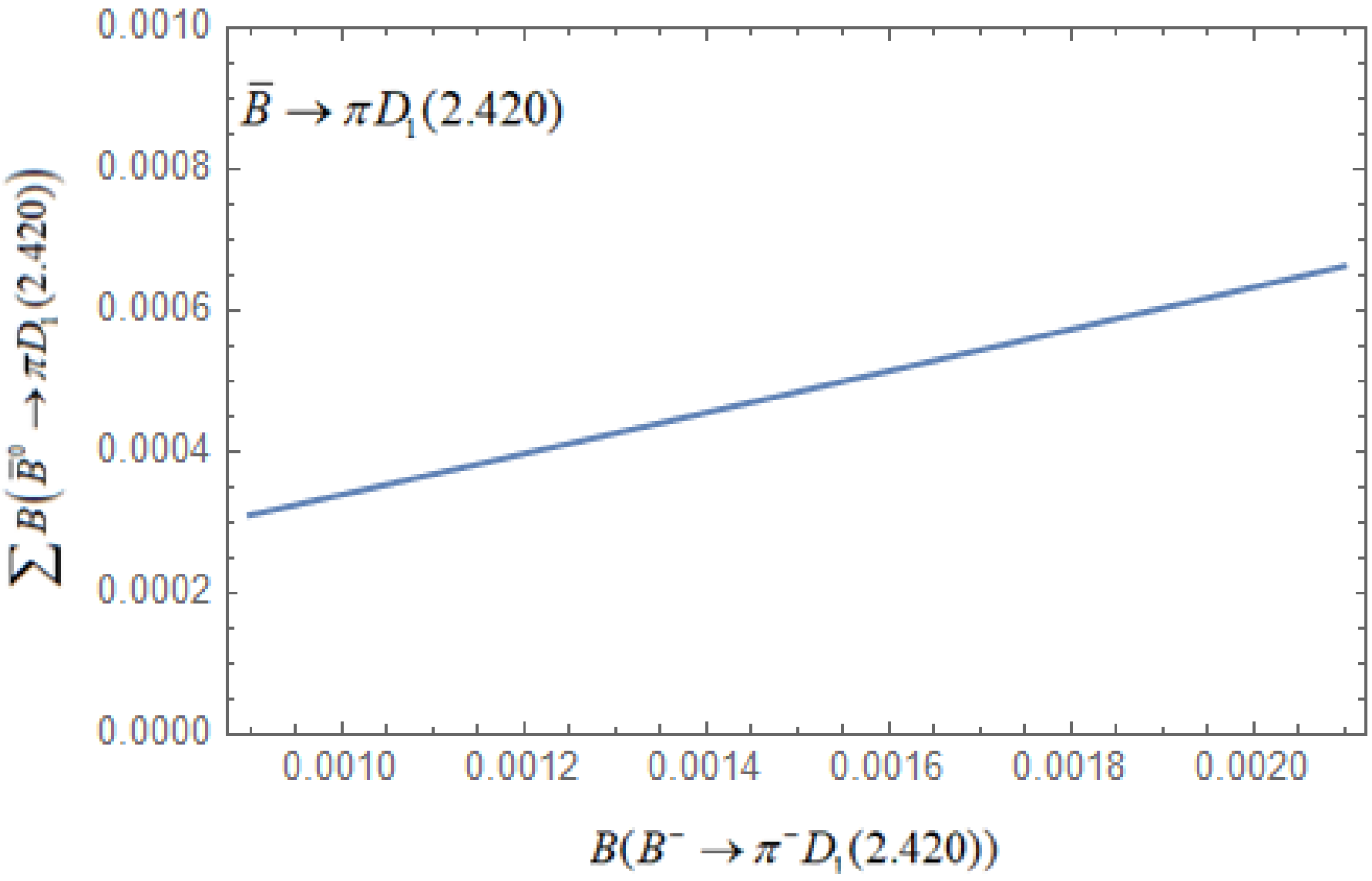

**Fig 2. Variation of** $\sum B\left(\bar{B}^0 \to \pi\ D_1\right)$ **with** $B(B^- \to \pi^-\ D_1^0)$

**A(iii)** $\bar{B} \to \pi D_1^{'}(2.427)$ **Decay mode**

The generic formula for $B^- \to \pi^- D_1^{'}(2.427)^0$ takes the following form:

$$B(\bar{B}^0 \to \pi^- D_1^{'+}) + B(\bar{B}^0 \to \pi^0 D_1^{'0})$$
$$= \frac{\tau_{\bar{B}^0}}{3\tau_{B^-}} B(B^- \to \pi^- D_1^{'0}) \times \left[ 1 + \left\{ \alpha + \frac{\left(\sqrt{2} - \alpha\right) A^f(\bar{B}^0 \to \pi^- D_1^{'+}) - \left(1 + \sqrt{2}\alpha\right) A^f(\bar{B}^0 \to \pi^0 D_1^{'0})}{A(B^- \to \pi^- D_1^{'0})} \right\}^2 \right]. \tag{62}$$

Here also, we take $\alpha \equiv \frac{A_{1/2}^{nf}}{A_{3/2}^{nf}} = 0.22$.

Using experimental branching $B(B^- \to \pi^- D_1^{'}(2.427)^0) \times B(D_1^{'}(2.427)^0 \to \pi^- D^{*-})$ given in Table 1, assuming that the $D_1^{'0}$ width is saturated by $\pi D^*$ [52], and then using isospin sum rule,

$$B(B^- \to \pi^- D_1^{'}(2.427)^0) \times B(D_1^{'}(2.427)^0 \to \pi^- D^{*+}) = 2/3, \tag{63}$$

we obtain $B(B^- \to \pi^- D_1^{'0}) = (7.5 \pm 1.7) \times 10^{-4}$, which yields

$$A(B^- \to \pi^- D_1^{'0}) = 0.213\,\text{GeV}^2. \tag{64}$$

Now we obtain factorizable amplitudes for $\overline{B}^0$ – decays, as given by

$$A^f(\overline{B}^0 \to \pi^- D_1^{'+}) = 2a_1 m_{D_1^{'}} f_\pi V_0^{\overline{B}D_1^{'3/2}}\left(m_\pi^2\right) = 0.106\,\text{GeV}^2.$$

$$A^f(\overline{B}^0 \to \pi^0 D_1^{'0}) = -\frac{1}{\sqrt{2}} 2a_2 m_{D_1^{'}} f_{D_1^{'3/2}} F_1^{\overline{B}\pi}\left(m_{D_1^{'3/2}}^2\right) = -0.029\ \text{GeV}^2. \tag{65}$$

Here,

$$f_{D_1^{'}} = -f_{D_1^{1/2}} \sin\theta_1 + f_{D_1^{3/2}} \cos\theta_1, \tag{66}$$

$$V_0^{\overline{B}D_1^{'}}\left(m_\pi^2\right) = -V_0^{\overline{B}D_1^{1/2}}\left(m_\pi^2\right)\sin\theta_1 + V_0^{\overline{B}D_1^{3/2}}\left(m_\pi^2\right)\cos\theta_1. \tag{67}$$

We use numerical values for the decay constants and form-factors, which have been presented in the last case. Finally, we predict

$$\sum B\left(\overline{B}^0 \to \pi D_1^{'}\right) \equiv B(\overline{B}^0 \to \pi^- D_1^{'+}) + B(\overline{B}^0 \to \pi^0 D_1^{'0}) == \begin{cases} (8.8 \pm 0.4) \times 10^{-4} \ \textit{for}\ \theta_1 = -8.1^\circ \\ (8.4 \pm 0.4) \times 10^{-4} \ \textit{for}\ \theta_1 = -3.3^\circ \end{cases} \tag{68}$$

Considering the uncertainty of the experimental branching $B(B^- \to \pi^- D_1^{'}(2.427)^0)$, here also, we plot the variation of the $\sum B\left(\overline{B}^0 \to \pi D_1^{'}\right)$ w.r.t $B(B^- \to \pi^- D_1^{'0})$ in Fig 3.

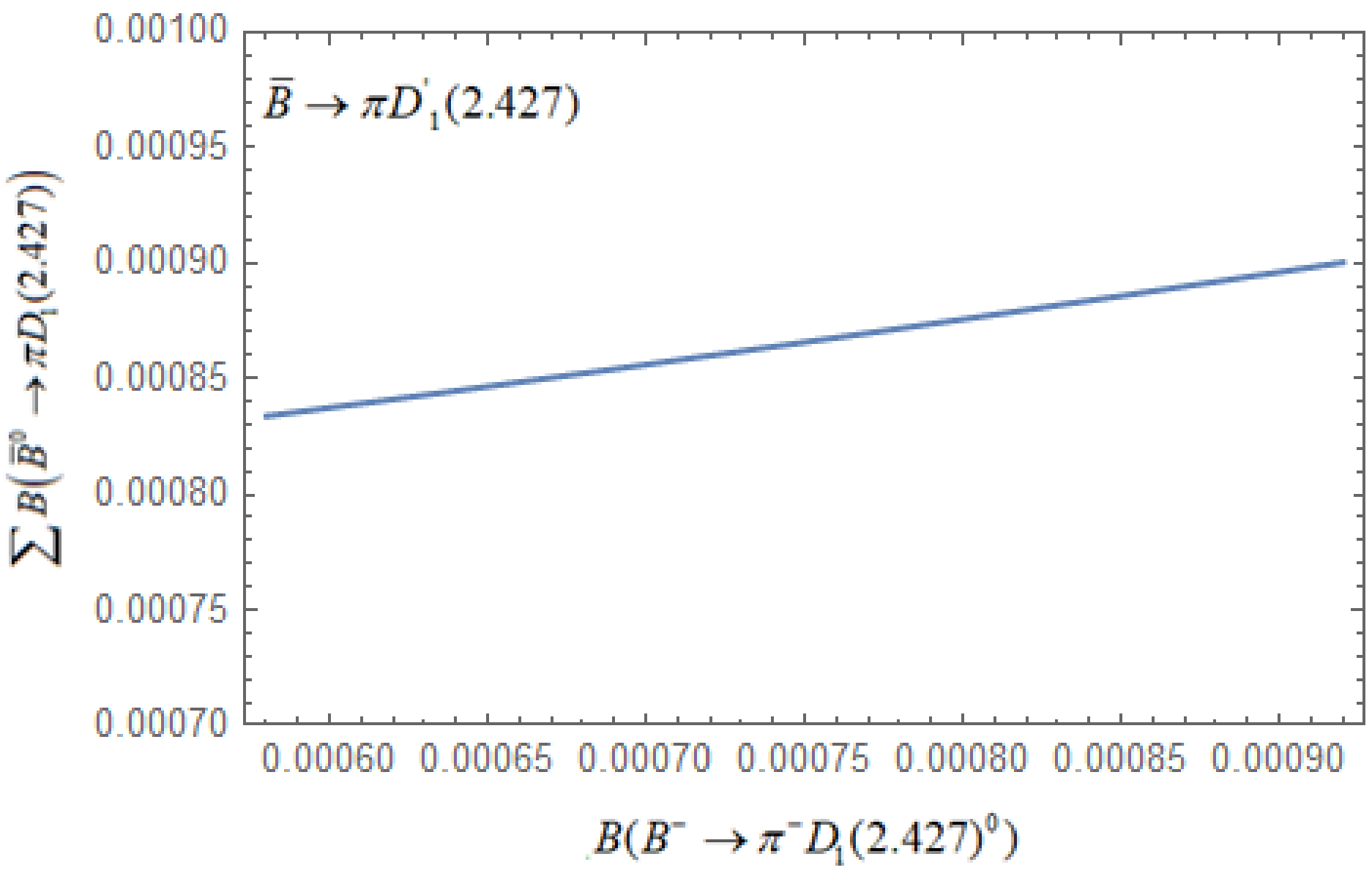

**Fig 3. Variation of** $\sum B\left(\bar{B}^0 \to \pi\, D_1^{'}\right)$ **with** $B(B^- \to \pi^-\, D_1^{'\ 0})$

B. $\bar{B} \to \pi D^*_2$ **Decay mode**

Experimentally [1], the tensor meson sixteen-plet comprises of isovector $a_2(1.320)$, strange iso-spinor $K_2^*(1.430)$, charm triplet $D_2^*(2.460)$, $D_{s2}^*(2.573)$, and three isoscalars $f_2(1.270)$, $f_2^{'}(1.525)$, and $\chi_{c2}(1P)$. These states behave well with respect to quark model assignments. For $\bar{B} \to PT$ decays, only one mode has been observed [1], $B(B^- \to \pi^- D_2^{*0}(2.460))$, and more data is expected to come in near future.

Writing the generic formula for $\bar{B} \to \pi D^*_2$ decays

$$B(\bar{B}^0 \to \pi^- D^{*+}_2) + B(\bar{B}^0 \to \pi^0 D^{*0}_2)$$
$$= \frac{\tau_{\bar{B}^0}}{3\tau_{B^-}} B(B^- \to \pi^- D^{*0}_2) \times \left[ 1 + \left\{ \alpha + \frac{\left(\sqrt{2}-\alpha\right) A^f(\bar{B}^0 \to \pi^- D^{*+}_2) - \left(1+\sqrt{2}\alpha\right) A^f(\bar{B}^0 \to \pi^0 D^{*0}_2)}{A(B^- \to \pi^- D^{*0}_2)} \right\}^2 \right]. \tag{69}$$

We proceed to calculate various quantities on the R.H.S. We combine both the results given in Table 1, *i.e.,*

$$B(B^- \to \pi^- D_2^*(2.462)^0) \times B(D_2^*(2.462)^0 \to \pi^- D^+) = (3.56 \pm 0.24) \times 10^{-4}, \tag{70}$$

$$B(B^- \to \pi^- D^*_2(2.462)^0) \times B(D^*_2(2.462)^0 \to \pi^- D^{*-}) = (2.2 \pm 1.0) \times 10^{-4}, \tag{71}$$

to arrive at

$$B(B^- \to \pi^- D_2^*(2.462)^0) \times B(D_2^*(2.462)^0 \to \pi^- D^+, \pi^- D^{*+}) = (5.7 \pm 1.1) \times 10^{-4}. \qquad (72)$$

Using $B(D_2^*(2.462)^0 \to \pi^- D^+, \pi^- D^{*+})$ =2/3 following from the isospin symmetry and assuming that the $D_2^{*0}$ width is saturated by $\pi D$ and $\pi D^*$ [52-61], we are led to

$$B(B^- \to \pi^- D_2^{*0}(2.462)) = (8.6 \pm 1.7) \times 10^{-4}. \qquad (73)$$

Using the branching fraction formula,

$$B\left(\bar{B} \to PT\right) = \tau_B \left| \frac{G_F}{\sqrt{2}} V_{cb} V_{ud}^* \right|^2 \frac{m_B^2 p^5}{12\pi m_T^4} \left| A\left(\bar{B} \to PT\right) \right|^2, \qquad (74)$$

where $p$ is the magnitude of the three-momentum of the final-state particle in the rest frame of $B$- meson and $m_B$ and $m_T$ denote masses of the $B$- meson and tensor meson, respectively.

By using the experimental value (72), we get

$$A(B^- \to \pi^- D_2^{*0}) = (6.5 \pm 0.6) \times 10^{-2} \text{ GeV}. \qquad (75)$$

The factorization parts of the weak decay amplitudes for $\bar{B} \to PT$ decays are expressed as the product of matrix elements of weak currents (up to the weak scale factor of $\frac{G_F}{\sqrt{2}} \times CKM\ elements \times QCD\ factors$ ),

$$\left\langle PT \middle| H_w \middle| B \right\rangle = \left\langle P \middle| J^\mu \middle| 0 \right\rangle \left\langle T \middle| J_\mu \middle| B \right\rangle + \left\langle T \middle| J^\mu \middle| 0 \right\rangle \left\langle P \middle| J_\mu \middle| B \right\rangle. \qquad (76)$$

The matrix elements $\left\langle P \middle| J^\mu \middle| 0 \right\rangle$ and $\left\langle P \middle| J_\mu \middle| B \right\rangle$ are given below. The hadronic current creating meson from the vacuum is given by

$$\left\langle P \middle| J^\mu \middle| 0 \right\rangle = i f_B P_B, \qquad (77)$$

where $P_B$ is the four- momentum of the pseudoscalar meson. However, the matrix elements $\left\langle T \middle| J^\mu \middle| 0 \right\rangle$ vanish due to the tracelessness of the polarization tensor $\varepsilon_{\mu\nu}$ of spin 2 meson and the auxiliary condition $q^\mu \varepsilon_{\mu\nu} = 0$ [62]. So, the tensor meson cannot be produced from the V-A current. Relevant $B \to T$ matrix elements are expressed as:

$$\begin{aligned} \left\langle T(P_T) \middle| J_\mu \middle| B(P_B) \right\rangle = {} & i\, h\, \varepsilon^{*\mu\nu} P_{B\alpha} (P_B + P_T)^\lambda (P_B - P_T)^\rho + k\, \varepsilon_{\mu\nu}{}^* P_B{}^\nu \\ & + b_+ (\varepsilon_{\alpha\beta}{}^* P_B{}^\alpha P_B{}^\beta) [(P_B + P_T)_\mu + b_- (P_B - P_T)_\mu], \end{aligned} \qquad (78)$$

in the ISGW2 model [5]. The matrix elements simplify to,

$$A(\overline{B} \to PT) = -i\, f_P\, F^{\overline{B}T}(m_P^2), \tag{79}$$

where

$$F^{\overline{B}T}(m_P^2) = k\,(m_P^2) + (m_B^2 - m_T^2)\, b_+(m_P^2) + m_P^2\, b_-(m_P^2). \tag{80}$$

Now we obtain factorizable amplitude values for $\overline{B}^0 -$ decays,

$$A^f(\overline{B}^0 \to \pi^- D_2^{*+}) = a_1 f_\pi F^{\overline{B}D^*_2}\left(m_\pi^2\right) = \begin{cases} 0.070 \; for\; F^{\overline{B}D^*_2}\left(m_\pi^2\right) = 0.52, \\ 0.051 \; for\; F^{\overline{B}D^*_2}\left(m_\pi^2\right) = 0.38, \end{cases} \tag{81}$$

using the decay constant values $f_\pi = -\left(0.131 \pm 0.002\right)$ GeV is already used in the previous sections [42], and the form factor $F^{\overline{B}D_2}\left(m_\pi^2\right) = 0.52,\, 0.38,$ taken from CLFQM [42] and ISGW model[3].

$$A^f(\overline{B}^0 \to \pi^0 D_2^{*0}) = -\frac{1}{\sqrt{2}}\, a_2 f_{D_2^*}\, F^{\overline{B}\pi}\left(m_{D_2^*}^2\right) = 0, \tag{82}$$

$A^f(\overline{B}^0 \to \pi^0 D_2^{*0})$ becomes zero due to vanishing of the decay constant of $D_2^*$ meson. Finally using (69) for $\alpha = 0.22$, we predict

$$B(\overline{B}^0 \to \pi^- D_2^{*+}) + B(\overline{B}^0 \to \pi^0 D_2^{*0}) = \begin{cases} (5.7 \pm 0.4) \times 10^{-4} \; for\; F^{\overline{B}D^*_2}\left(m_\pi^2\right) = 0.52\,; \\ (4.1 \pm 0.4) \times 10^{-4} \; for\; F^{\overline{B}D^*_2}\left(m_\pi^2\right) = 0.38\,; \end{cases} \tag{83}$$

for the two choices of $F^{\overline{B}D_2}\left(m_\pi^2\right) = 0.52,\, 0.38,$ respectively, which may be tested in future experiments. Considering the ambiguity of the experimental $B(B^- \to \pi^- D_2^*(2.460)^0),$ we find the increasing behavior of $\sum B\left(\overline{B}^0 \to \pi D_2\right)$ w.r.t $B(B^- \to \pi^- D_2^{*0})$ in Fig 4 for both the choices shown as dashed and thick lines, respectively.

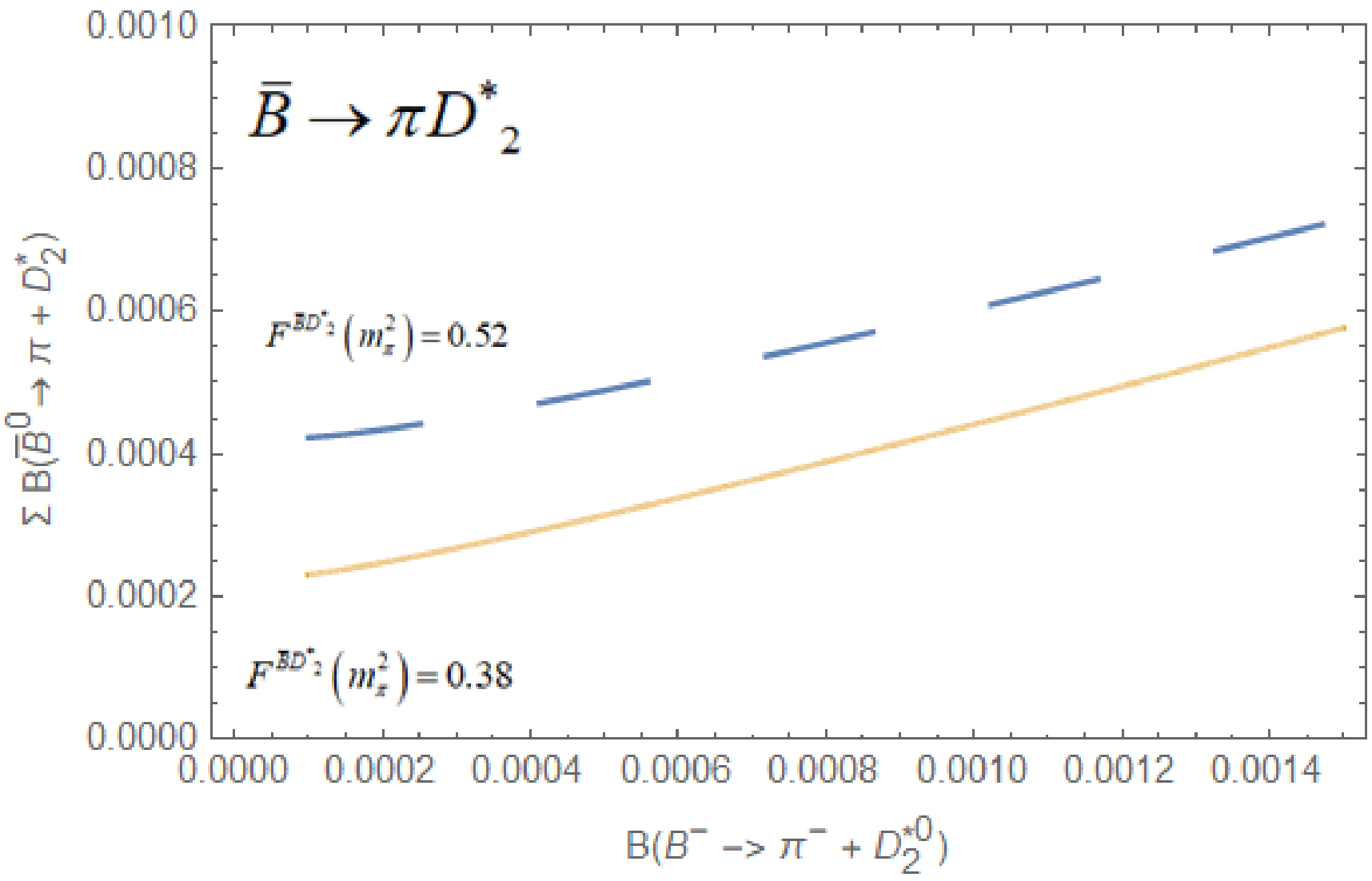

**Fig 4. Variation of $\sum B\left(\bar{B}^0 \to \pi\, D_2^*\right)$ with $B(B^- \to \pi^-\, D_2^{*0})$ for different values of form-factor**

### C. $\bar{B} \to \pi D_0^*$ Decay mode

The scalar mesons mostly appear as the hadronic resonances, and have large decay widths. There will exist several resonances and decay channels within a short mass interval. The overlaps between resonances and background make it considerably difficult to resolve the scalar mesons. The scalar-meson family has been the most difficult one to identify as a standard sixteen-plet. Experimentally [1], the following states of scalar meson sixteen-plet, isovector $a_0(0.980)$, strange spinor $K_0^*(1.429)$, one isoscalar $\chi_{c0}(1P)(3.145)$, charm triplet $D_0^*(2.400), D_{s0}^*(2.480)$, behave well with respect to quark model assignments. For $\bar{B} \to PS$ decays, only one mode has been observed [1], $B(B^- \to \pi^- D_0^{*0}(2.400))$, and more data is expectedto come in near future.

Writing the generic formula explicitly for $\bar{B} \to \pi D_0^*$ decays,

$$B(\bar{B}^0 \to \pi^- D_0^{*+}) + B(\bar{B}^0 \to \pi^0 D_0^{*0})$$
$$= \frac{\tau_{\bar{B}^0}}{3\tau_{B^-}} B(B^- \to \pi^- D_0^{*0}) \times \left[ 1 + \left\{ \alpha + \frac{\left(\sqrt{2} - \alpha\right) A^f(\bar{B}^0 \to \pi^- D_0^{*+}) - \left(1 + \sqrt{2}\alpha\right) A^f(\bar{B}^0 \to \pi^0 D^{*0}_0)}{A(B^- \to \pi^- D_0^{*0})} \right\}^2 \right]. \tag{84}$$

In order to obtain the branching fraction $B(B^- \to \pi^- D_0^{*0})$ from the experimental value

$$B(B^- \to \pi^- D_0^*(2.400)^0) \times B(D_0^*(2.400)^0 \to \pi^- D^+) = (6.4 \pm 1.4) \times 10^{-4}, \tag{85}$$

given in Table 1, we employ isospin symmetry which gives.

$$\frac{\Gamma(D_0^{*0} \to \pi^- D^+)}{\Gamma(D_0^{*0} \to \pi^0 D^0) + \Gamma(D_0^{*0} \to \pi^- D^+)} = \frac{2}{3}, \tag{86}$$

and realizing the saturation of strong $D_0^{*0}$ decays with $D_0^{*0} \to \pi\ D$ modes [52], we estimate

$$B(B^- \to \pi^- D_0^*(2.400)^0) = (9.6 \pm 2.1) \times 10^{-4}, \tag{87}$$

for our analysis. Using this estimate and the decay rate formula, similar to that of $\overline{B} \to PP$,

$$B\left(\overline{B} \to PS\right) = \tau_B \left|\frac{G_F}{\sqrt{2}} V_{cb} V_{ud}^*\right|^2 \frac{p}{8\pi m_B^2} \left|A\left(\overline{B} \to PS\right)\right|^2, \tag{88}$$

we get

$$A(B^- \to \pi^- D_0^{*0}) = \left(1.06 \pm 0.32\right) \times 10^{-4}\ \text{GeV}^3. \tag{89}$$

We then obtain factorizable amplitudes for $\overline{B}^0$ – decays, which are given as:

$$\begin{aligned} A^f(\overline{B}^0 \to \pi^- D_0^{*+}) &= a_1 f_\pi (m_B^2 - m_{D^*_0}^2) F^{\overline{B}D^*_0}\left(m_\pi^2\right) = 0.824\,\text{GeV}^3, \\ A^f(\overline{B}^0 \to \pi^0 D_0^{*0}) &= -\frac{1}{\sqrt{2}} a_2 f_{D^*_0} (m_B^2 - m_\pi^2) F^{\overline{B}\pi}\left(m_{D^*_0}^2\right) = -0.0522\ \text{GeV}^3 \end{aligned} \tag{90}$$

Numerical values are calculated using the decay constants [42],

$$f_\pi = \left(0.131 \pm 0.002\right)\ \text{GeV}, \quad f_{D_0^*} = \left(0.107 \pm 0.013\right)\ \text{GeV}. \tag{91}$$

and $F^{\overline{B}D^*_0}(m_\pi^2)$ from the CLFQM [42] results, *i.e.,*

$$F^{\overline{B}D_0^*}\left(q^2\right) = \frac{F^{\overline{B}D_0^*}(0)}{\left(1 - a\left(\frac{q^2}{m_B^2}\right) - b\left(\frac{q^2}{m_B^2}\right)^2\right)}, \tag{92}$$

where

$$\begin{aligned} F^{\overline{B}D_0^*}\left(0\right) &= (0.27 \pm 0.01), \\ a = 1.08 \pm 0.04, &\quad b = 0.23 \pm 0.02, \end{aligned} \tag{93}$$

and the form-factor $F^{\overline{B}\pi}(0) = 0.27 \pm 0.05$ is already given in previous sections.

Finally, we predict

$$\sum B\left(\bar{B}^0 \to \pi D_0^*\right) \equiv B(\bar{B}^0 \to \pi^- D_0^{*+}) + B(\bar{B} \to \pi^0 D_0^{*0}) = (4.8 \pm 0.6) \times 10^{-4}, \qquad (94)$$

for $\alpha = 0.22$. Here also, we plot the variation of the $\sum B\left(\bar{B}^0 \to \pi D_0^*\right)$ w.r.t $B(B^- \to \pi^- D_0^{*0})$ in Fig 5, which also shows the increasing behaviour.

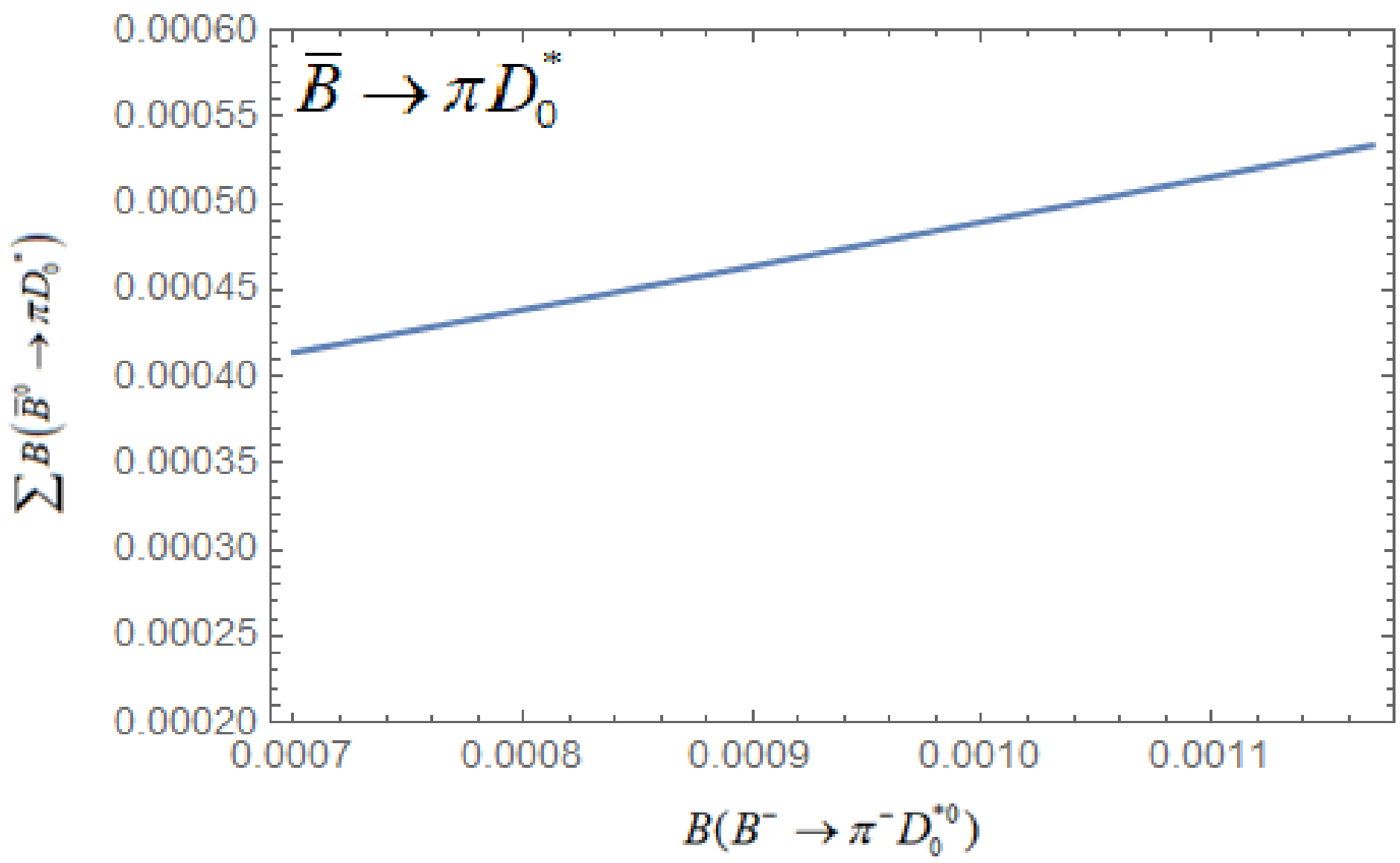

**Fig 5. Variation of** $\sum B\left(\bar{B}^0 \to \pi\ D_0^*\right)$ **with** $B(B^- \to \pi^-\ D_0^{*0})$

## V. SUMMARY AND CONCLUSIONS

In our previous work, we have carried out isospin analysis of CKM-favored two-body weak decays of bottom mesons $\bar{B} \to PP/PV$, occurring through *W*-emission quark diagrams. Obtaining the factorizable contributions from the spectator-quark model for $N_c$ = 3 (real value), we have determined nonfactorizable reduced isospin amplitudes from the experimental data for these modes. We have observed that in all the decay modes, the nonfactorizable isospin reduced amplitude $A_{1/2}^{nf}$ bears the same ratio with $A_{3/2}^{nf}$ within the experimental errors. In the charm sector a systematics observed for the charm mesons decaying to *s*-wave mesons has been found to be consistent with their *p*-wave meson emitting decays [23]. Encouraged by the success for the *s*-wave emitting decays in bottom meson sector [37], we have extended isospin analysis for the *p*- wave meson emitting decays in $\bar{B} \to PA/PT/PS$ channels, particularly for the $\bar{B} \to a_1 D/\pi D_1/\pi D_1^{'}/\pi D_2/\pi D_0$ decays, which have the same isospin structure as that of $\bar{B} \to \pi D/\rho D/\pi D^*$ cases.

In order to include the effects of nonfactorizable contributions, for these cases, we exploit the generic formula to predict sum of the branching fractions of $\bar{B}^0-$decays in these channels. As there are large errors involved in $B(\bar{B}\to a_1 D)=(4\pm 4)\times 10^{-3}$ and the form-factor $F^{\bar{B}a_1}(0)$ is not uniquely known, looking at these uncertainties, we plot the variation of $\sum B\left(\bar{B}^0\to decays\right)$ *w.r.t* $B(B^-\to a_1^- D^0)$ for extreme values of $V_0^{\bar{B}a_1}(0)=0.14$ and 1.20, which enhances our prediction by a factor of 1.19. Our predictions will be tested in the future experiments.

We extend our analysis to $\bar{B}\to \pi D_1/\pi D_1^{'}/\pi D_2/\pi D_0$ decay modes, which have the similar isospin structure and make predictions for $\bar{B}^0-$decays. It is hoped that the predictions made in this paper would help the experimentalists to identify the *p*- wave meson emitting decays of the heaviest bottom mesons.

## VI. REFERENCES

[1] P. A. Zyla *et al.* (Particle Data Group), "Review of Particle Physics", Prog. Theor. Exp. Phys, **2020**: 083C01 (2020)

[2] M. Wirbel, B. Stech and M. Bauer, "Exclusive Semileptonic Decays of Heavy Mesons", Z. Phys. C, **29**: 637 (1985)

[3] M. Bauer, B. Stech and M. Wirbel, "Exclusive Nonleptonic Decays of D, D(s), and B Mesons", Z. Phys. C, **34**:103 (1987)

[4] M. Wirbel, "Description of weak decays of D and B mesons", Prog. Part. Nucl. Phys, **21**: 33-48 (1988)

[5] N. Isgur, D. Scora, B. Grinstein and M. B. Wise, "Semileptonic B and D decays in the quark model", Phys. Rev. D, **39**: 799 (1989)

[6] D. Scora and N. Isgur, "Semileptonic meson decays in the quark model: An update", Phys. Rev. D, **52**: 2783 (1995)

[7] T. E. Browder and K. Honscheid, "B mesons", Prog. Nucl. Part. Phys, **35**: 81-255 (1995)

[8] Shuo-Ying Yu and Xian-Wei Kang, " Nature of X(3872) from its radiative decay", Phys. Lett. B, **848**: 138404 (2024)

[9] H. Y. Cheng, "Effects of final-state interactions in hadronic B decays", Int. J. Mod. Phys. A, **2**: 650 (2006)

[10] M. Suzuki, "The final-state interaction in the two-body nonleptonic decay of a

heavy particle", [hep-ph/9807414].

[11] Wei Wang and Rui-Lin Zhu,"To understand the rare decay $B_s \to \pi^+\pi^- l^+ l^-$ ", Phys. Lett. B, **743**: 467-471 (2015)

[12] Wen-Fei Wang, Hsiang-nan Li, Wei Wang and Cai-Dian L¨u "The S-wave resonance contributions in the $B^0{}_s \to J/\Psi\pi^+\pi^-$ and $B_s \to \pi^+\pi^-\mu^+\mu^-$ decays", Phys. Rev. D, **91**: 094024 (2015)

[13] Yu-Ji Shi and Wei Wang, "Chiral Dynamics and S-wave contributions in Semileptonic $D_s/B_s$ decays into $\pi^+\pi^-$", Phys. Rev. D, **92**: 074038 (2015)

[14] Yu-Ji Shi, Wei Wang and Shuai Zhao, "Chiral dynamics, S-wave contributions and angular analysis in $D\to\pi\pi\ell\bar{\nu}$", Eur. Phys. J. C, **77**: 452 (2017)

[15] M. Doring, Ulf-G. Meißner and Wei Wang, "Chiral dynamics and S-wave contributions in semileptonic B decays", J. High Energ. Phys., **2013**: 11 (2013)

[16] Wei Wang, "Search for the a0(980)–f0(980) mixing in weak decays of Ds/Bs mesons", Phys. Lett. B, **759:** 501–506 (2016)

[17] M. Doring, Ulf-G. Meißner, Wei Wang, "Chiral Dynamics and S-wave Contributions in Semileptonic B decays", JHEP, **10:** 011 (2013)

[18] Yu-Ji Shi, Chien-Yeah Seng, Feng-Kun Guo, Bastian Kubis, Ulf-G. Meißner and Wei Wang, "Two-meson form factors in unitarized chiral perturbation theory", JHEP, **04:** 086 (2021)

[19] Wei Wang, "On the production of hidden-flavored hadronic states at high energy", Chin. Phys. C, **42:** 043103 (2018)

[20] Ye Xing and Zhi-Peng Xing, "S-wave contributions to in the perturbative QCD framework", Chin. Phys. C, **43**: 7, 073103 (2019)

[21] R. C. Verma, "SU (3) flavor analysis of nonfactorizable contributions to D→ PP decays", Phys. Lett. B, **365**: 377 (1996)

[22] R. C. Verma, "Searching a systematics for nonfactorizable contributions to hadronic decays of $D^0$ and $D^+$ mesons", Z. Phys. C, **69**: 253 (1996)

[23] A. C. Katoch, K. K. Sharma and R. C. Verma, "Isospin analysis of non-factorizable contributions to hadronic decays of charm mesons", J. Phys. G, **23**: 807 (1997)

[24] A. N. Kamal, A. B. Santra, T. Uppal and R. C. Verma, " Nonfactorization in Hadronic two-body Cabibbo favored decays of $D^0$ and $D^+$", Phys. Rev. D, **53**: 2506 (1996)

[25] C. K. Chua, W. S. Hou and K. C. Yang, " Final state rescattering and Color-suppressed $\overline{B}^0 \to D^{(*)0}h^0$ decays", Phys. Rev. D, **65**: 096007 (2002)

[26] C. K. Chua and W. S. Hou, " Rescattering effects in $\overline{B}_{u,d,s}{}^0 \to DP, \overline{D}P$ decays", Phys. Rev. D, **77**: 116001 (2008)

[27] C. W. Chiang and E. Senaha, " Update analysis of two-body charmed B meson decays", Phys. Rev. D, **75:** 074021 (2007)

[28] S. H. Zhou, Y. B. Wei, Q. Qin, Y. Li, F. S. Yu and C. D. Lu, "Analysis of two-body charmed B meson decays in factorization- Assisted Topological- Amplitude Approach", Phys. Rev. D, **92:** 094016 (2015)

[29] L. Szymanowski, "QCD collinear factorization, its extension and partronic distribution", arxiv:1208:5688

[30] H. n. Li, Y. L. Shen, and Y. M. Wang, " Next-to-leading-order corrections to B→π form factors in kT factorization", Phys. Rev. D, **85**: 074004 (2012)

[31] S. Cheng, Y. Y. Fan, X. Yu, C. D. Lu, and Z. J. Xiao, "The NLO twist-3 contributions to B→π form factors in kT factorization", Phys. Rev. D, **89**: 094004 (2014)

[32] N. Deshpande, M. Gronau and D. Sutherlard, "Final State Gluon Effects in Charmed Meson Decays", Phys. Lett. B, **90**: 431 (1980)

[33] M. Shifman, "Factorization versus duality in weak nonleptonic decays", Nucl. Phys. B, **388**: 346 (1992)

[34] A. N. Kamal and T. N. Pham, "Cabibbo favored hadronic two-body B decays", Phys. Rev. D, **50**: 395 (1994)

[35] A. J. Buras, J. M. Gerard and R. Ruckl, "1/n Expansion for Exclusive and Inclusive Charm Decays", Nucl. Phys. B, **268:** 16-48 (1986)

[36] H. Y. Cheng and B. Tseng, "Nonfactorizable effects in spectator and penguin amplitudes of hadronic charmless B decays", Phys. Rev. D, **58:** 094005 (1998)

[37] Maninder Kaur, Supreet Pal Singh and R. C. Verma, "Searching a systematics for nonfactorizable contributions to $B^-$ and $\overline{B}^0$ mesons", Chin. Phys. C, **46:**

073105 (2022)

[38] G. Buchalla, A. J. Buras and m. E. Lautenbacher, “Weak deacys beyond leading logarthims”, Rev. Mod. Phys, **68**: 1125 (1996)

[39] S. H. Zhou, Y. B. Wei, Q. Qin, Y. Li, F. S. Yu and C. D. Lu, “Analysis of two-body charmed B meson decays in factorization- Assisted Topological- Amplitude Approach”, Phys. Rev. D, **92**: 094016 (2015)

[40] S. H. Zhou, R. H. Li, Z. Y. Wei and C. D. Lü, “Analysis of three-body charmed B meson decays under the factorization-assisted topological-amplitude approach”, Phys. Rev. D, **104:** 116012 (2021)

[41] M. Wirbel, “Description of weak decays of D and B mesons”, Prog. Part. Nucl. Phys, **21**: 33-48 (1988)

[42] R. C. Verma, “Decay constants and form factors of s-wave and p-wave mesons in the covariant light-front quark model”, J. Phys. G, **39**: 025005 (2012)

[43] A. K. Giri, L. Maharana and R. Mohanta, “Two body non-leptonic decays $B \to D\pi$ with heavy quark and chiral symmetry”, Pramana – J. Phys, **48**: 5 (1997)

[44] H. Na, et al., HPQCD Collaboration, “ $B \to Dl\nu_l$ form factors at nonzero recoil and extraction of $|V_{cb}|$**”,** Phys. Rev. D, **92:** 054510 (2015), Erratum: Phys. Rev. D, **93:** 119906 (2016)

[45] R. C. Longacre, “The E(1420) meson as a $KK\bar{\pi}$ molecule”, Phys. Rev. D, **42:** 874 (1990)

[46] F. Aceti, J.-J. Xie and E. Oset, “The KK¯π decay of the f1(1285) and its nature as a $K\bar{K}\pi$ cc molecule”, Phys. Lett. B, **750**: 609 (2015)

[47] Yu-Ji Shi, Jun Zeng and Zhi-Fu Deng, “Revisiting $K_1(1270) - K_1(1400)$ mixing with QCD sum rules”, Phys. Rev. D, **109**: 016027 (2024)

[48] Huseyin Dag, Atug Ozpineci and Ayce Cagil and Guray Erkol, “ Theoretical determination of $K_1(1270,1400)$ mixing J. Phys. Conf. Ser, **348**: 012012 (2012)

[49] Zhi-Qing Zhang, Hongxia Guo and Si-Yang Wang, “ Study of the $K_1(1270) - K_1(1400)$ mixing in the decays $B \to J/\psi K_1(1270), J/\psi K_1(1400)$ ”, Eur. Phys. J. C, **78**: 3, 219 (2018)

[50] M. Suzuki, “Strange axial-vector mesons”, Phys. Rev. D, **47**: 1252 (1993)

[51] G. Claderon et al., "Nonleptonic two-body B-decays including axial-vector mesons in the final state", Phys. Rev. D, **79**: 094019 (2007)

[52] K. Abe *et al.,* "Study of $B^{-} \rightarrow D^{**0}\pi^{-}(D^{**0} \rightarrow D^{*+}\pi^{-})$ decays", (Belle Collaboration), Phys. Rev. D, **69**: 112002 (2004)

[53] P. Colangelo *et al.,* "Bounding effective parameters in the chiral Lagrangian for excited heavy mesons", Phys. Lett. B, **634**: 235 (2006)

[54] A. Deandrea, R. Gatto, G. Nardulli and A. D. Polosa, "Semileptonic $B \rightarrow \rho$ and $B \rightarrow a_1$ transitions in a quark-meson model", Phys. Rev. D, **59**: 074012 (1999)

[55] T. M. Aliev and M. Savci, "Semileptonic $B \rightarrow a_1 l\nu$ decays in QCD", Phys. Lett. B, **456**: 256 (1999)

[56] K. C. Yang, " Form factors of $B_{u,d,s}$ Decays into P-Wave Axial-Vector Mesons in the Light-Cone Sum Rule Approach", Phys. Rev. D, **78**: 034018 (2008)

[57] R. H. Li, C. D. Lu and W. Wang, "Transition form factors of B decays into p-wave axial- vector mesons in the perturbative QCD approach", Phys. Rev. D, **79**: 034014 (2009)

[58] Pallavi Gupta and A. Upadhyay, "Analysis of strong decays of charmed mesons $D_2^*(2460), D_0(2560), D_2(2740), D_1(3000), D_2^*(3000)$ and their spin patners $D_1^*(2680), D_3^*(2760)$ and $D_0^*(3000)$ ", Phys. Rev. D, **97**: 1, 014015 (2018)

[59] H. Y. Cheng and C. K. Chua and C. W. Hwang, "Covariant light front approach for s wave and p wave mesons: Its application to decay constants and form-factors", Phys. Rev. D, **69:** 074025 (2004)

[60] H. Y. Cheng, "Implications of Recent $\overline{B}^0 \rightarrow D^{*0}X^0$ Measurements", Phys. Rev. D, **65**: 094012 (2002)

[61] H. Y. Cheng, "Phenomenological applications of QCD factorization to semi-inclusive B decays", [hep-ph/0109259]

[62] A. C. Katoch and R. C. Verma, "Weak decays of B- and anti-B0 mesons to a Pseudoscalar meson and a tensor meson involving a b ---> c transition", Phys. Rev. D, **79**: 1717 (1995)